\documentclass[twocolumn]{article}
\usepackage{fancyhdr}
\usepackage{graphicx}
\usepackage{epstopdf}
\usepackage{chronology}
\usepackage[superscript]{cite}
\usepackage{caption}
\usepackage[version=3]{mhchem}
\usepackage{color}
\usepackage[titles]{tocloft}
\begin{document}

\twocolumn[{

\title{\bf \textsf{Multiple Exciton Generation in 
Nanostructures\\ for Advanced Photovoltaic Cells}}
\author{Nicholas Siemons, Alessio Serafini}
\maketitle

  \begin{@twocolumnfalse}
\begin{abstract}
{This paper reviews both experimental and theoretical work on nanostructures showing high quantum yields due to the phenomenon of multiple exciton generation. It outlines the aims and barriers to progress in identifying further such nanostructures, and also includes developments concerning solar devices where nanostructures act as the light-absorbing component. It reports on both semiconductor and carbon structures, both monocomposite (of various dimensionalities) and heterogeneous. Finally, it looks at future directions that can be taken to push solar cell efficiency above the classic limit set by Shockley and Queissier in 1961.}
\end{abstract}
\vspace{0.8cm}
  \end{@twocolumnfalse}
}]

\tableofcontents

\parskip = 3mm

\section{Introduction}

\begin{figure*}
  \begin{center}
\scalebox{0.4}{\includegraphics{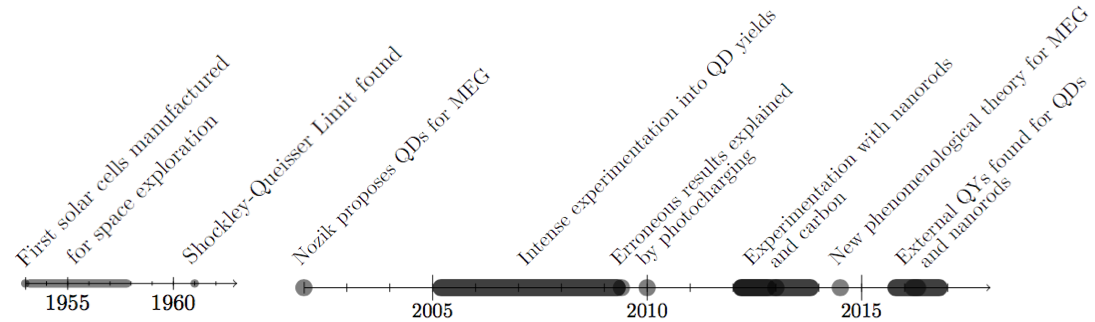}} 
\caption{Timeline showing the main contributions to understanding and implementation of multiple exciton generation.
\vspace{0.7cm}} \label{timeline}
  \end{center} 
  \hrulefill
\end{figure*}

\begin{figure}[t!]
  \begin{center}
\scalebox{0.3}{\includegraphics{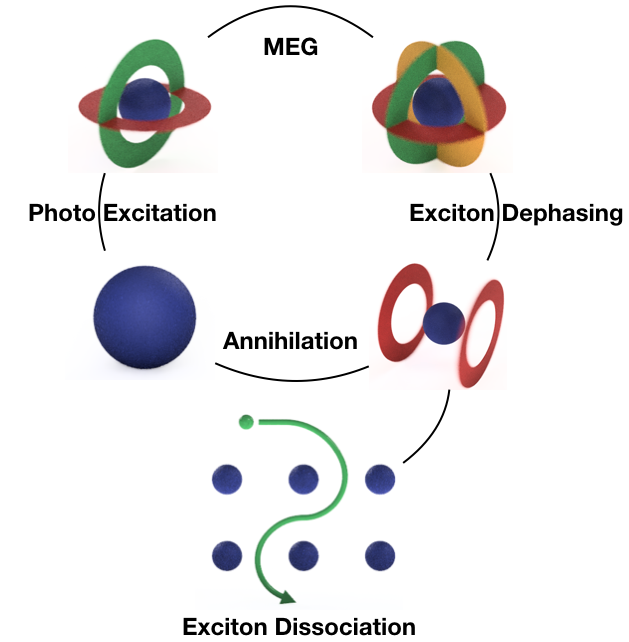}} 
\caption{Multiple Exciton Generation in a photovoltaic device, adapted from  \cite[H. Jaeger]{Jaeger2013}.} \label{meg2}
  \end{center}
  \hrulefill
\end{figure}

\begin{figure}[b!]
\hrulefill
  \begin{center}
\scalebox{0.31}{\includegraphics{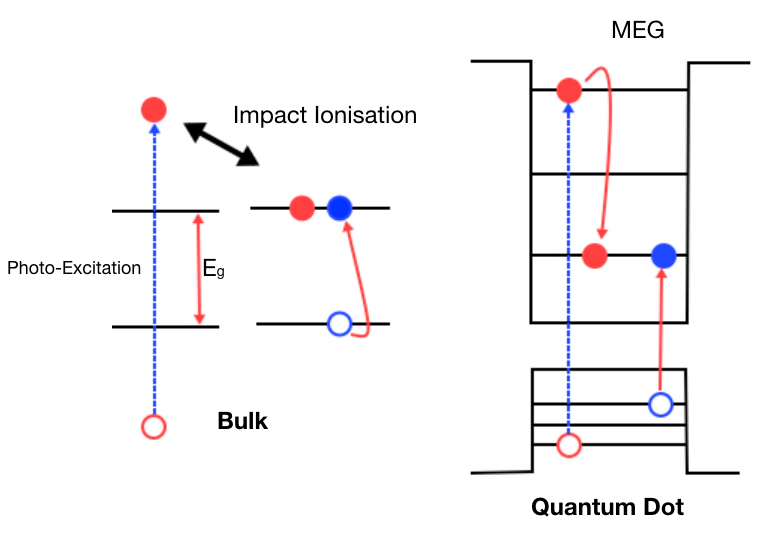}} 
\caption{Generation of multiple excitons from highly excited electron-hole pairs. {\bf Left}: in bulk semiconductor, were the excited electron relaxes to the band gap.  {\bf Right} in QD's impact ionisation is the inverse of the Auger process, and the rate of this is much larger than in bulk. It is usually referred to as MEG in QDs. Adapted from \cite[A. Shabaev]{Shabaev2013}.} \label{meg3}
  \end{center}
\end{figure}

Current solar energy conversion inefficiency is considered by many to be the greatest barrier to sustainably meeting modern day energy needs. Since the early 20\textsuperscript{th} century there has been increasing interest in solar voltaics, with early models of p-n junction solar cells manufactured since the 1950's (see figure \ref{timeline}).  In 1961 William Shockley and Hans Queisser discovered a fundamental limit to the efficiency of a traditional p-n solar cell of 30\% \cite{Shockley1961}.  At the end of the 20th century second generation solar cells were manufactured which use multiple layers of semiconducting material. These devices absorb a greater range of frequencies and do surpass this limit, pushing the theoretical limit to a 68\% efficiency. However, the manufacture of such multi-junction solar cells is too expensive for large scale commercial use \cite{C0JM01178A}.  
Theory and experiments show there is potential to overcome the Shockley-Queisser limit  using quantum effects for a new cheap third generation solar cell \cite{Scully2010, Krogstrup2013} (figure \ref{meg2}). In fact, it is known that impact ionisation in bulk semiconductor materials allows for a hot excitons to relax to the valence band, in turn exciting another exciton (figure \ref{meg3}). However, in bulk, this does not occur at the typical frequencies of solar photons. Therefore, intensive research is being done to the aim of increasing the quantum yield (QY) of photovoltaic cells by enhancing biexciton creation in nanostructures, which allow for a significant response to a wider frequency spectrum. Such structures include nanosheets, or ``platelets'', nanorods, nanowires and, notably, quantum dots (QDs). In nanostructures, electron-hole pair multiplication, whereby more than one (typically two) electron-hole pairs are generated by impact with a single incoming photon, is referred to as multiple exciton generation (MEG), to differentiate it from bulk impact ionisation. Let us mention that alternative terminologies, such as `carrier multiplication' (CM) or `multiple exciton collection' (MEC) are also in use in the literature. It is also worth knowing that 0-dimensional structures (quantum dots) may also be referred to as ``nanocrystals''.

This review paper is organised as follows: in Sec.~\ref{secmeg}, the notion of multiple exciton generation 
is cursorily recalled, along with some basic related terminology;  Sec.~\ref{sec0d} covers MEG in zero-dimensional (quantum dots) structures, with an emphasis on \ce{Pb} and \ce{Si} based dots;
Sec.~\ref{sec1d} moves on to one-dimensional structures, in particular to nanorods, nanotubes and nanoribbons; Sec.~\ref{sec2d} is devoted to two-dimensional structures, such as generic nanosheets and graphene sheets; Sec.~\ref{sechetero} completes our survey focusing on hybrid type II heterostructures, which have drawn interest in this context; finally, in Sec.~\ref{secsum}, we include a summary and a few conclusive remarks.


\section{Multiple Exciton Generation}\label{secmeg}

The creation of additional carriers from hot excitons has been observed in bulk semiconductors such as \ce{PbS},\ce{PbSe} and \ce{Si} \cite{Robbins1980}, and was not included in the seminal analysis by Shockley and Queisser. {Theoretical work by De Vos {\it et. al.} has however shown that a QY higher than unity is beneficial for increasing solar device efficiency \cite{Badescu2001,De.Vos1998,DeVos1993,Gent1995}}. In a p-n junction solar cell, it has been shown that efficient impact ionisation occurs when the incident photon energy is in excess of five times the energy $E_g$ required to excite an exciton to the valence band \cite{Beard2010}.  Typically, this is high energy ultra-violet light outside the solar spectrum. Hence, a key aim in order to exploit this effect is to reduce the threshold energy $\hbar \omega_{th}$ for it to take place, so that quantum yields (i.e., the number of excitons per incoming photon) greater than unity can be reached with solar photons. {Before 2002, MEG was often referred to as `down-shifting' or `down-conversion' of hot excitons, and devices were theorized which would facilitate this process \cite{Trupke2002,Badescu2007}.} In 2002 however, Nozik was the first to propose that MEG could be enhanced by using nanostructures \cite{nozik2002quantum}. Research published soon after by Shaller and Klimov showed evidence for low threshold MEG in \ce{PbSe} QDs \cite{PhysRevLett.92.186601}.
Since then, MEG has been observed repeatedly in a variety of QDs such as \ce{PbSe}\cite{C0JM01178A}, \ce{PbS}\cite{Schaller2006} and \ce{Si}\cite{Beard2007}, while some semiconductor QDs such as \ce{InAs} have as of yet shown no signs of MEG \cite{Nair2007, Pijpers2007}.  Notably, MEG has also been demonstrated in \ce{C} structures, such as nanotubes, nanoribbons and graphene \cite{Gabor2009}. 

It is worth noticing that such studies are typically 
carried out in a colloidal environment, using either photoluminescence, transient absorption or, more recently, pump-probe spectroscopy.  All of these methods are labour intensive and difficult and, indeed, the lack of quick and easy MEG tests is arguably one of the major barriers in the field \cite{Gao2015}.  As a consequence, it is only recently that nanostructures {have been put into actual devices. The study of such devices is arguably a new field in itself, involving practical and 
technical considerations that elude the main scope of this review. 
Although we shall mention several of the most important advances concerning devices, the reader is referred 
to \cite[C. Stolle]{Stolle2013} for a more detailed and exhaustive account.}

The early reported QYs of experiments (2005-1010) varied widely up to extremes of 300\% \cite{Beard2007, Ellingson2005, Beard2011, doi:10.1021/nl100177c}.  It was later discovered that some suspiciously optimistic results were due to photo-charging, whereby electrons delocalise from parent QDs.  This gives a deceptive MEG-like signature in the spectroscopic data. Subsequently, it was found that this effect could be nullified simply by stirring the sample solution \cite{Midgett2010}. Another hurdle in the assessment of QYs for specific structures is clearly the large number of variables involved and relevant to these effects, 
such as size, composition, shape and method of observation. It is uncommon for studies to be done on like-for-like samples, so that corroborative results are rare \cite{B913277P}. The complicated interplay between the phonon cooling rate and the nanostructure dimensions is also an important reason for the wide variety in reported QY values \cite{Kumar2016}.
In fact, there is still no general, unified theoretical framework for describing MEG. Nevertheless, all of the studies in the area agree on two key factors: the enhancement of the Coulombic coupling between exciton and biexciton states and the suppression of excitons' radiative relaxation pathways (predominantly phonon emission) are critical for increasing QY \cite{Jaeger2013, Shabaev2013, Piryatinski2010, LeBris2010}. Such conditions can be achieved by reducing the size of the basic structures, as quantum confinement favours both of them. 

As far as the general theory of MEG is concerned, it should be noted that the intriguing possibility of mechanisms hinging 
on coherent superpositions of multiple excitonic states has been put forward, but it does not agree with all experiments \cite{Ellingson2005, doi:10.1021/nl062059v}.  More recently `window of opportunity' models have been used \cite{Piryatinski2007, Stewart2013}, which allow for more predictive power than previous modelling.  They rely on splitting relaxation times into discrete amounts, and are often complemented with computer simulations. The main aim in these studies is identifying conditions whereby the threshold photon energy for multiple excitation, $\hbar \omega_{th}$, and the electron-hole pair energy $\epsilon_{eh}$ (the energy required for a further exciton to be created) are minimised.  In the ideal case scenario, they would approach the limit set by the conservation of energy, when $\hbar \omega_{th} = 2E_g$ and $\epsilon_{eh} = E_g$, $E_g$ being the band-gap energy. This produces a staircase function when plotting pump energy against QY, which is often taken as an optimal reference. 

{It should be mentioned that, when discussing in-situ structures, the observed QY is sometimes also termed ``quantum efficiency''. When dealing with devices, however, one is typically only interested in their quantum efficiency. This is further qualified as either `external' or `internal' quantum efficiency.  The former is the ratio of generated electrons to {\it incident} photons, whilst the latter is the ratio of electrons to {\it absorbed} photons, which takes into account reflection and transmission, and is always lower. Furthermore, 
let us remark that, in understanding the relationship between in-situ structures and devices, the exciton lifetime $\tau$ plays a key role, as a larger $\tau$ allows for excitons to be `collected', thus increasing the device's quantum efficiency.}


\section{Quantum Dots}\label{sec0d}

Due to their comparative topological simplicity and ease of fabrication, QDs are the best understood class of nanostructures. It is thus not surprising that efforts have been made to design effective ways to assemble devices from QDs and for turning excitons into carriers \cite{Lan2014}. 

\begin{figure}
  \begin{center}
\scalebox{0.41}{\includegraphics{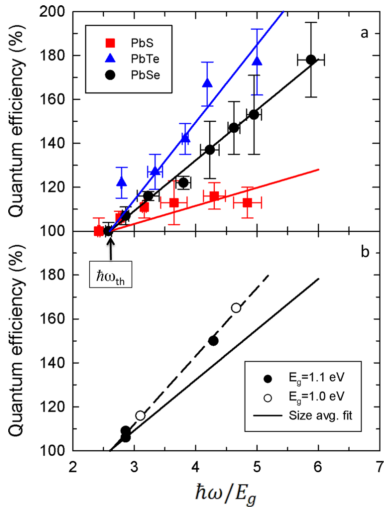}} 
\caption{An example of MEG for three most studied \ce{PbX} QDs as a function of standardised pump energy, $\hbar \omega / E_g$.  In a you can see significantly different QYs, indicative of different $\epsilon_{eh}$ for each QD.  From \cite[L. Padilha]{1Padilha2013}. } \label{pbx}
  \end{center} 
  \hrulefill
\end{figure}

In order to be competitive candidates for solar voltaic devices, QDs must have $E_g$ within the infrared region, high QY, and be stable in the cell environment. For these reasons, \ce{PbS}, \ce{PbSe}, \ce{PbTe}, \ce{Si} and \ce{TiO2} QDs have received the most attention (figure \ref{pbx}).  Computational methods are used in order to identify further possible options \cite{Jaeger2013, RevModPhys.86.153}. Typically, such methods adopt density-functional theory, which provides a satisfactory compromise between accuracy and scale. In \ce{Si29H36}, the MEG dynamics in QDs was probed using density-functional theory \cite{2016APS..MARH24002M}, placing the threshold for efficient MEG at about $3E_g$ which, for \ce{Si29H36}, lies within the visible spectrum. Such computational techniques are bound to furnish additional insight as methods and machines improve. On the other hand, computational methods are typically not designed explicitly to resolve the excited states occupied by biexcitons, which poses challenges and limitations to this approach.

As pointed out above, it is important that photovoltaic structures exhibit quantum confinement. In this regard, theory suggests that for this to be the case the radius of the QD should be less than the bulk exciton Bohr radius, calculated as $a_B = \epsilon \hbar^2 / \mu e^2$, with $\epsilon$ and $\mu$ being, respectively, the dielectric constant and the reduced mass of electron and hole.  Typical values for such Bohr radii are $a_B(\ce{PbSe}) = 46 nm$, $a_B(\ce{PbS}) = 20 nm$ and $a_B(\ce{Si}) = 4.7 nm$ \cite{BEARD2014}.  QDs of these sizes can now by synthesised easily.

\subsection{\ce{PbX} Quantum Dots}

Lead based QDs have been studied intensively, as here $E_g$ is in the low infra-red spectrum, fabrication is easily scaleable, reproducible and low-cost and, most importantly, the Bohr-exciton radius tends to be large. After initial studies in 2009 it was doubted that \ce{PbSe} QDs, which, it should be mentioned, feature the added complication of being unstable in air, would exhibit MEG at all \cite{PhysRevLett.92.186601,Pijpers2009}. However, in 2010 Beard \textit{et al.} \cite{Beard2010} showed sufficient evidence for MEG, and now \ce{PbSe} QDs are possibly the most studied structures in this arena. 
In 2014, an improvement in \ce{PbSe} synthesis was introduced that allowed the QDs to survive for up to thirty days in air, showing an increase of orders of magnitude over previous lifetimes \cite{Zhang2014}. this was made possible by a novel synthetic method whereby \ce{PbSe} QDs are made directly from \ce{CdSe} QDs by displacing the \ce{Cd} with \ce{Pb}, creating a protective \ce{Cl} and \ce{Cd} ion surface on the subsequent \ce{PbSe} QD film (in order to check that the other characteristics of the QD films were the same as standard ones direct comparisons were carried out, showing that the MEG dynamics are unchanged).
\begin{center}
\begin{table}[b!]
    \begin{tabular}{  p{4cm}  p{1cm}  p{1cm}  p{1cm}  }
    \hline

\bf{Pump } & {\bf QY} & & \\ 
\bf{Wavelength (nm)} & $\text{QD}_1$ & $\text{QD}_2$ & $\text{QD}_3$ \\ \hline 
800 & 1 & 1 & 1 \\ 
400 & 1.10 & 1.16 & 1.09 \\ 
267 & 2.19 & 2.29 & 2.38 \\
    \hline    
    \end{tabular}
    \caption{An example of recent data collected for QD QYs for three sizes of \ce{PbSe} QDs from \cite{Kumar2016}.  $r(\text{QD}_1) = 1.8 \pm 0.2 nm$, $r(\text{QD}_2) = 1.9 \pm 0.3 nm$, $r(\text{QD}_3) = 2.2 \pm 0.3 nm$ in diameter.}
    \label{pbse}
    \end{table}
        \end{center}   
In a 2016 study the QY of PbSe QDs of three different sizes were determined, see Table \ref{pbse}\cite{Kumar2016}, for radii of $1.8 \pm 0.2 \, nm$, $1.9 \pm 0.3 \, nm$ and $2.2 \pm 0.3 \, nm$ respectively -- all much smaller than the \ce{PbSe} Bohr radius, and thus in the strong quantum confinement region. The peaks in transition absorption observed in the experiment may be associated to specific, previously known orbital energies \cite{Koole2008}.  In particular, the so-called $\Sigma$ transitions to the `Brillouin Zone' (unique to \ce{PbSe}), a set of electronic transitions where MEG is expected to be particularly strong due to an enhanced phonon bottleneck, were investigated by 
probing the biexciton dynamics through transient absorption. The
measures of the QY of each QD at three different pump energies thus highlighted the central role played by electron excitation to the $\Sigma$ levels in increasing efficiency, in agreement with the 
related theory. 
\begin{figure*}
  \begin{center}
\scalebox{0.6}{\includegraphics{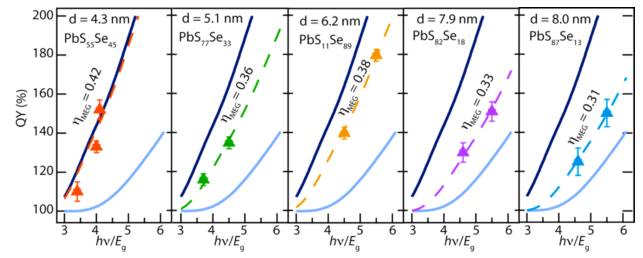}} 
\caption{QYs for different sizes of \ce{PbS_xSe_{1-x}} alloy QDs from \cite[A. Midgett]{Midgett2013} plotted against the standardised pump energy.  The solid dark line is showing the QYs for \ce{PbSe} QDs, while the light blue line is for bulk \ce{PbS}.  The dotted lines are least-squares best fit lines for the alloy dots.} \label{pbsse2}
  \end{center}
  \hrulefill
\end{figure*}

\begin{figure}
  \begin{center}
\scalebox{0.35}{\includegraphics{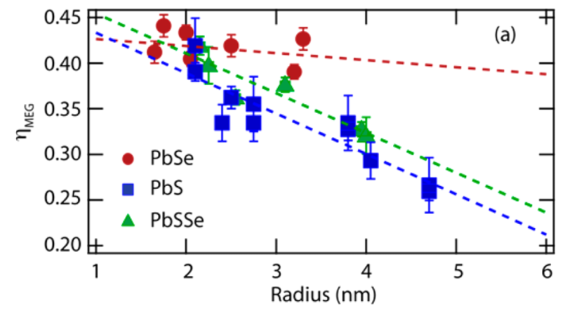}} 
\caption{An example of MEG efficiency data plotted against the radius of \ce{PbS}, \ce{PbSe} and \ce{PbS_xSe_{1-x}} from \cite[A. Midgett]{Midgett2013}.} \label{pbsse}
  \end{center}
  \hrulefill
\end{figure}

\ce{PbX} QDs share similar crystal structure, and so systematic, comparative investigations may be launched where the second element is changed, in order to isolate the impact of different variables on the overall yields \cite{Midgett2013}.  A 2011 study made the comparison between \ce{PbSe} and \ce{PbS} to determine why QY for \ce{PbSe} seemed much higher \cite{Stewart2012}.  These two QD's are studied particularly as \ce{PbS} QDs are stable in air and have the same structure as \ce{PbSe} QDs which, as already mentioned, are relatively unstable in air.  By looking at excited-state lifetimes it was established that the exciton-biexciton Coulomb coupling was very similar in both QDs.  After observing a dependancy of QY on the QD size for \ce{PbS} but not for \ce{PbSe}, it was also possible to infer that other relaxation channels, specifically the phonon relaxation rate, must be the dominant factor at play in determining the different QYs. This result was instrumental in leading to the first medium-scale collection of QY data on a variety of lead-based QDs of different sizes \cite{Midgett2013} (Figure \ref{pbsse2}). In this paper, the accompanying theory is  refined with the introduction of the parameters $\eta$, an MEG efficiency value independent of the pump rate, and $P$, a factor which describes the competition between the rate of MEG ($\kappa_{\text{\tiny{MEG}}}$) and the rate of non-useful cooling ($\kappa_{\text{\tiny{cool}}}$).  In \ce{PbSe}, MEG occurs at about $2.8E_g$, which is described by a $\eta$ value between 0.4 and 0.5 (Figure \ref{pbsse}).  The following empirical equation relates $P$ with the photon energies involved:
\begin{equation}
\frac{\kappa_{\text{\tiny{MEG}}}}{\kappa_{\text{\tiny{cool}}}} = P \cdot \Big(\frac{\hbar \nu_{ex}}{\hbar \nu_{th}} \Big)^s
\end{equation}
where $s$ is a factor which describes how MEG changes with  $\hbar \nu_{th}$ and the excess pump photon energy $\hbar \nu_{ex}$.  Here, $s$ is found to be around 2.2 for lead based QDs.  The study by Midgett \textit{et al.}, in agreement with previous relevant literature \cite{Midgett2010, 2016APS..MARH24002M}, also reports on properties for a variety of \ce{PbS_xSe_{1-x}} alloy QDs, with varying composition and radius, allowing one to draw conclusions about the size-dependance of $\eta$ for the different compositions of QDs (see figure \ref{pbsse}). It is suggested that higher MEG might be achievable in non-spherical crystal structures, although more data needs to be collected to explore such configurations.

\subsection{\ce{Si} Quantum Dots}

Silicon is abundant in nature. It makes up over 90\% of solar voltaic devices currently in use and plays a huge role in electronics and technology. It poses no environmental issues and \ce{Si} membranes are optically transparent to non-bandgap energy photons, such that it would be possible to create \ce{Si} multilayer cells of different sizes in order to span a wider absorption spectrum and achieve greater efficiency \cite{Hao2009}. It also has QD $E_g$ within the infra-red making it potentially an ideal material for ${\rm 3}^{{\rm rd}}$ generation solar cells.  

QY above unity with silicon QDs was first reported in 2007 by Beard \textit{et al.} \cite{Beard2007}, who tested QDs of two different diameters, one in the strong confinement region (below the Bohr exciton radius for \ce{Si} of $4.9 \, nm$) and one in the weak-intermediate confinement region (about double the Bohr exciton radius). MEG was thus observed in a colloidal environment for both of these QDs at a pump energy of $0.86 \, eV.$  The reconstructed value for the electron-hole energy was $\epsilon_{eh} = (2.4 \pm 0.1)E_g$ (close to the energy-conservation limit of $2E_g$), with a reported QY of $2.6\pm0.2$ at $3.4E_g$.  Other relaxation pathways other than phonon emission were also taken into account in this study, such as charge transfer and radiative recombination \cite{Scully2010}.  

In 2016, Bergren \textit{et al.} used pump-probe spectroscopy to study a range of differently sized \ce{Si} QDs \cite{Bergren2016}. As will be discussed below, in recent years emphasis has been on trying to extend biexciton lifetimes, which is the angle taken by Bergren and coworkers in this work. Thus they found that, as one should expect, biexciton lifetimes increase as the diameter is reduced, and postulate that the increase in the quantum confinement is reducing the coupling to the phonon relaxation channels.   This is consistent with evidence for other QDs.  

Study into \ce{Si} QD films is also intensive due to easy access to high-grade samples.  In  \cite[M. Trinh]{Trinh2012}, a novel way to induce MEG in an array of \ce{Si} QDs was described, showing that if the QDs are in close proximity, hot excitons can excite excitons in neighbouring QDs rather than in their parent QD.  Specifically, this experiment used \ce{Si} QDs of around $1.8 \, nm$ embedded in an \ce{SiO2} matrix. MEG obtained through such a thermal exciton propagation enjoys a reduced chance of Auger recombination and and exciton lifetime prolonged by up to six orders of magnitude. A recent paper by Kryjevski {\it et. al.} \cite{Kryjevski2015a} adopts density functional theory to evaluate the effect of changing the internal structure of the QDs in amorphous QDs, as opposed to crystalline ones: such computer simulations predict a considerable change in dynamics in the 2-dimensional density limit present in films.  They suggest that in this case, amorphous QDs may show a marked improvement in QY, along with better carrier mobility.   

Recently, research has increasingly focused on novel structures rather than mere manipulation of the QD size.  The production and characterisation of multi-layer all \ce{Si} cells with different band gaps due to different levels of \ce{P} dopant, rather than due to different sizes, have been first explored \cite{Hao2009}.  High levels of MEG in \ce{GeX} QDs and higher QYs than in bulk \ce{GeX} were subsequently observed in these structures \cite{Saeed2015}, with {external quantum efficiency exceeding 120\% for \ce{PbTe} QDs \cite{Bohm2015}. It is perhaps worth noting that such studies,} which focus on {\em external} QYs, are key to showing that the in-situ studies feature behaviours can be replicated for 
real-world usage. 

\section{One-Dimensional Structures}\label{sec1d}

Over the last few years, one-dimensional structures have started to draw attention as candidates for photovoltaic devices alternative to quantum dots. Whilst clearly more challenging to synthesise than zero-dimensional dots, 
higher-dimensional structures entail a breaking of the spherical symmetry that lifts some of the energy level degeneracies and thus allows for additional excitations. Besides, theoretical models suggested that, despite the increased volume of 1-dimensional structures such as nanorods, they would still lie in the quantum confinement region, as the effect of quantum confinement depends not merely on the total volume of a structure but rather on its 
effective cross sectional area \cite{Bartnik2010, Li2001}.  In terms of MEG, nanowires have only been considered in preliminary computational studies \cite{Kryjevski2015, 2016APS..MARH24002M}. On the other hand cell architectures based on nanorods and nanotubes have received some attention \cite{Garnett2011}. 
\begin{figure}[b!]
\hrulefill
  \begin{center}
\scalebox{0.28}{\includegraphics{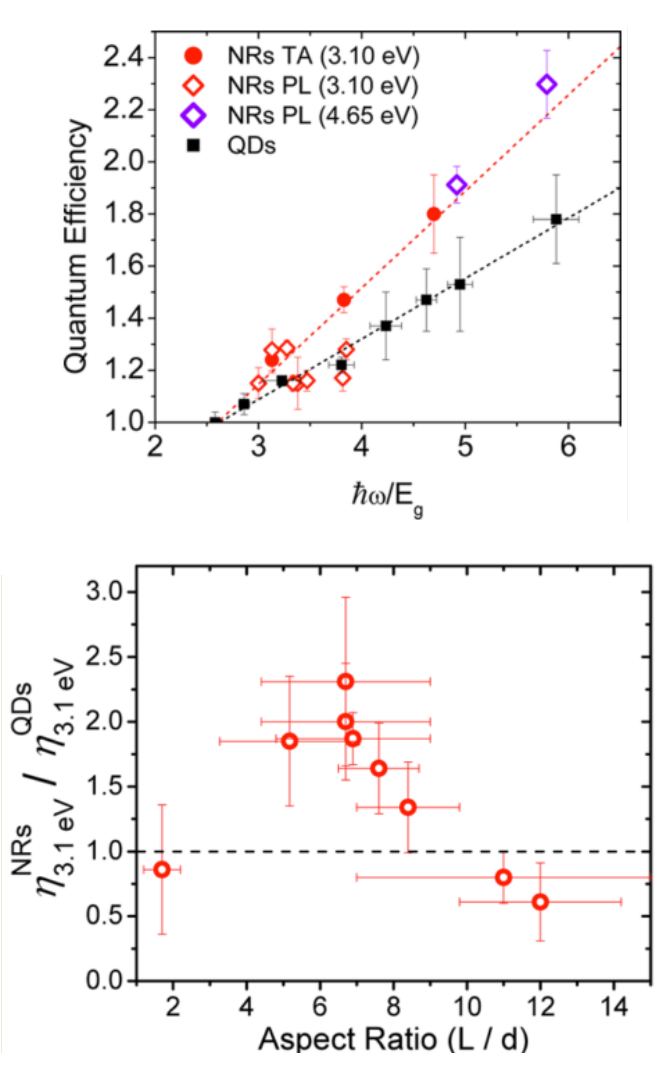}} 
\caption{From \cite[L. Padilha]{Padilha2013}.  {\bf Top}:   QYs measured for a range of nanorods of different aspect ratios and band gap energies compared against the same range of results for quantum dots.  The data for the quantum dots is from \cite[J. McGuire]{McGuire2008, McGuire2010}. {\bf Bottom}:  Carrier Multiplication enhancement factor for nanorods against QDs as a function of aspect ratio.  } \label{yoyo}
  \end{center} 
\end{figure}

\subsection{Nanorods}

\ce{Pb}-based nanorods were the first non-spherical structures to be studied for MEG, which was made possible 
by synthetic methods developed in 2010 \cite{koh2010, Wang2010}. Initial reports showed the first evidence of MEG \cite{Sandberg2012, Cunningham2011}. In particular, in a paper by Cunningham {\it et. al.}\cite{Cunningham2011} values of $\hbar \omega_{th}$ and $\epsilon_{eh}$ close to the ideal values of $2E_g$ and $E_g$ were reported. The key issue in these studies is clearly to achieve a general understanding the effect of length and aspect ratio on QYs.  
The \ce{PbSe} nanorods' QY dependance on aspect ratio (defined as $\rho = \text{length}/\text{diameter}$) was investigated soon afterwards \cite{Padilha2013,1Padilha2013}, indicating that higher efficiency may be due to increased exciton Coulombic couplings at certain $\rho$, as suggested in \cite[A. Bartnik]{Bartnik2010}.  After testing samples with different $\rho$ and $E_g$, it was found that the QY has a nonlinear, non-monotonic dependence with $\rho$, and $\eta$, displaying an almost threefold-increase at certain $\rho$ values - see Figure \ref{yoyo}; in particular, the efficiency peaks at $\rho = 6 - 7$ give a two-fold increase in QY over \ce{PbSe} QDs. The `enhancement factor' (defined as $\eta_{\text{nanorod}}/\eta_{\text{QD}}$ for the same composition), varies significantly across the literature. 
\begin{figure*}
  \begin{center}
\scalebox{0.27}{\includegraphics{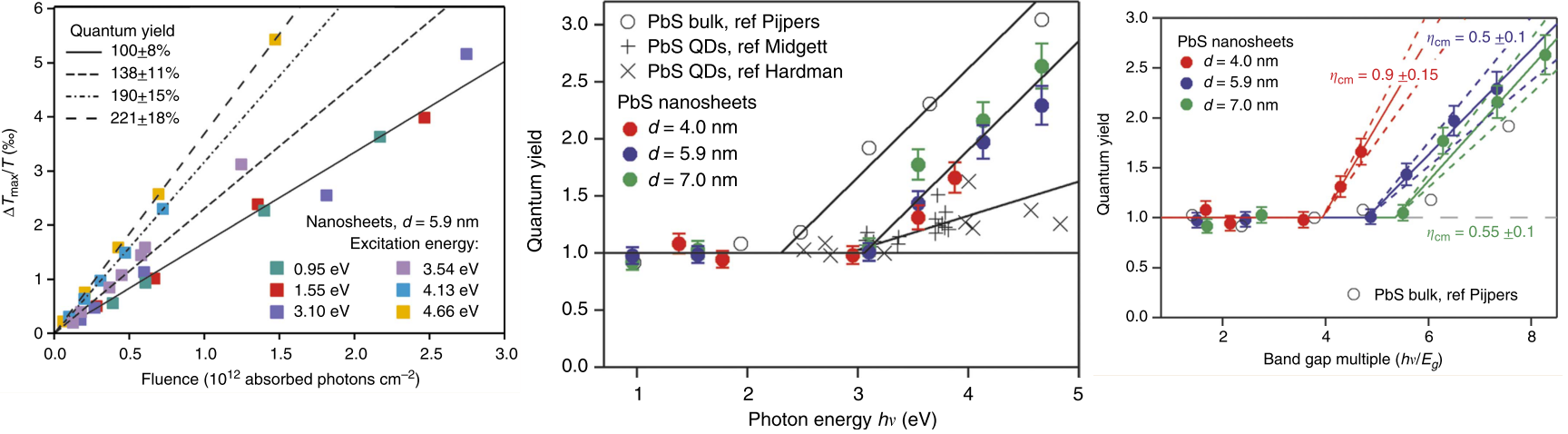}} 
\caption{{\bf Left}: Maximum bleach signal obtained from transient absorption, against pump photon energies for \ce{PbS} nanosheets of $5.9$nm thickness.  The solid line implies QY = 1, and the steeper lines imply higher QYs.  {\bf Middle and right}: Quantum yield against pump photon energy $\hbar \omega$, or standardised pump photon energies $\hbar \omega/E_g$.  Slope of data in middle and right figures is equal to $\eta$. Figures from  \cite[M. Aerts]{Aerts2014}.} \label{sheet1}
  \end{center} 
  \hrulefill
\end{figure*}
In late 2015, Davis \textit{et al.} measured an external quantum efficiency exceeding unity for \ce{PbSe} nanorods \cite{Davis2015}. More specifically, external QYs of $109 \pm 3 \%$, $113 \pm 3 \%$ and $122 \pm 3 \%$ 
were measured for devices containing nanorods of $E_g = 1.05$eV, $E_g = 0.95$eV and $E_g = 0.80$eV.  This is a promising result, which gives confidence that the same structures could be used in future devices. In the same study, 
the internal QY is shown to be in very good agreement with pre-existing experimental and theoretical literature \cite{Padilha2013, 1Cunningham2013}.

\subsection{Carbon Nanotubes and Nanoribbons}

Theoretical work from 2004 suggested strong Coulombic interactions between carriers would be present in  single-walled carbon nanotubes (SWCNTs) \cite{Spataru2004, Perebeinos2004}. This and other characteristics such as interesting photo and electric properties, as well as strong photon absorption, imply that such structures have the potential for MEG. Despite being considered a 1d material,  the dynamics that govern MEG in SWCNTs is different to nanorods or nanowires. This is because excitons are confined to the surface of the SWCNTs rather than at their centre, which implies a quasi-2d behaviour.  

An initial 2008 study tested a SWCNT sample with mixed chiralities \cite{Ueda2008} showing a QY of 130\% for $\hbar \omega = 3.7E_g$.  A paper published two years later performed the same experiment with SWCNTs of specific chiralities \cite{Wang2010a}.  The latter, by Wang \textit{et al.}, achieved the same QY at $\hbar \omega = 3E_g$ for (6,5) SWCNTs. However, both these results show values of $\hbar \omega_{th}$ outside the solar spectrum.  Biexciton creation was further reported in 2013 \cite{Yuma2013}, although no QY is quoted there. 
A basic hurdle toward the exploitation of SWCNTs in photovoltaic cells is represented by the difficulty of synthesing  samples of a single chirality, which require labour-intensive retro-processes.  For more information about the uses of SWCNTs in solar voltaic system outside of enhancing MEG, see \cite[D. Tune]{Tune2013} and \cite[D. Jariwala]{Jariwala2013}.

Recently, MEG has been investigated in graphene nano-ribbons (GNRs) using transient absorption \cite{Soavi2016}.  GNRs have the advantage of having synthesis methods of atomic precision \cite{Cai2010}. Furthermore, MEG turns out to be enhanced in these structures compared to SWCNTs, as they can be made extremely narrow with a significantly smaller surface area per unit length. The study of GNRs may still be considered to be in its infancy. However, initial theoretical reports indicate potential for extraordinary photo-electric properties \cite{Prezzi2011,Yang2007,Prezzi2007}.

\section{Nanosheets}\label{sec2d}

Synthetic methods for semiconductor nanosheets are still imprecise and difficult, and as such are the greatest barrier to their study \cite{Nasilowski2016}.  In 2014, Aerts \text{et al.} \cite{Aerts2014} used a method of synthesis described in \cite{Schliehe2010} to make ultra-thin \ce{PbS} nanosheets and test them via transient absorption (figure \ref{sheet1}). Sheets of three thickness, $4.0\pm0.1$, $5.9\pm0.1$ and $7.0\pm0.1$nm gave values for the maximum transient absorption bleaching  $\Delta T_{max}/T$ (measure of the proportion of absorbed photons, where $T$ is temperature), which could be inserted into the equation 
\begin{equation}
\frac{\Delta T_{max}}{T} = \Omega \, \sigma_{TB} \, I_0 \, F_A
\label{Eq5}
\end{equation}
to determine the QY, $\Omega$.  The transient bleach cross section $\sigma_{TB}$ was determined from low-energy pumping with a known incident number $I_0 \, F_A$ of photons per unit area.  The data analysis shows the notable result that $\hbar \omega_{th} = 3 \, eV$ is not affected by the nanosheet's thickness or $E_g$. Most remarkably, it also shows that, for the thinnest nanosheet of $4$nm, one has $\eta = 0.9\pm0.15$: This is extremely close to the optimal value of $\eta = 1$ and would suggest that all excess energy has gone into exciting multiple carriers. Note that this is significantly larger than the reported values for \ce{PbX} QDs, which is around $\eta = 0.3-0.5$ \cite{Pijpers2009, Midgett2013, Padilha2013}. Unfortunately, these nanosheets have the disadvantage of high $\hbar \omega_{th}$, although the ability to exhibit near-perfect efficiency once the threshold frequencies are reached is extremely interesting per se.

\subsection{Graphene}

A more readily available 2d structure with well established synthesis is graphene, whose promise for useful optoelectric properties, such as high carrier densities and charge mobility, is well known \cite{Bonaccorso2010}.  Between 2010 and 2012, theoretical work by Winzer \textit{et al.} showed that graphene has the potential for MEG \cite{Winzer2010, Winzer2012}.  Only a year later, Tielrooij {\em et al.} \cite{Tielrooij2013} carried out a study on 
MEG biexciton creation in graphene through methods similar to those that would then be used in \cite{Aerts2014} on \ce{PbSe} nanosheets. Such a study identifies two dominant processes: carrier-carrier scattering, leading to MEG, and phonon-assisted cooling, as well as a branching ratio between the two, which is shown to be highly favourable for MEG. This study employed doped graphene, and suggests a dependance of the results on the doping levels, which would open the additional possibility to alter the threshold energies for MEG by changing the amount of dopant as with \ce{Si}.  Soon after, in 2014, MEG was detected in low-doped graphene too \cite{Plotzing2014}.  

It should be mentioned that graphene has been studied in a wide variety of roles inside devices,  not only as the light-absorbing component.  To find out more about its role in enhancing photo-induced voltage in dye sensitised solar cells and QD cells, see the review paper \cite[C. Ubani]{Ubani2016}.

\section{Type II Heterostructures}\label{sechetero}

\begin{figure}
  \begin{center}
\scalebox{0.43}{\includegraphics{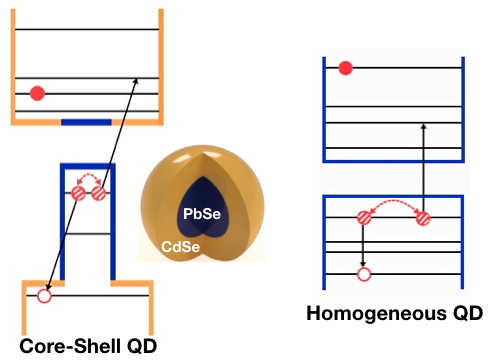}} 
\caption{Graphic of the energy levels and states in core-shell QDs. Adapted from \cite[C. Cirloganu]{Cirloganu2014}. 
} \label{het1}
  \end{center} 
  \hrulefill
\end{figure}

New QD structures that aim to separate the hole and electron wavefunctions have emerged recently. 
The wavefunction separation suppresses Auger Recombination and thus increases the exciton lifetime \cite{Piryatinski2007}.  These `type II' heterostructures are created by using multiple different compositions in the same structure (figure \ref{het1}). 

A prominent example of a heterostructure is represented by core-shell QDs, such as those studied in Cirloganu \textit{et al} in \cite{Cirloganu2014} on the basis of a phenomenological model for MEG proposed by John Stewart \cite{Stewart2013}, which suggests that slowing down hot carrier relaxation might be even more important than previously thought. In general, this work has spurred an interest in cooling-rate engineering for nanostructures, and in turn much research into how the separation between electron and hole wavefunctions may be increased. 

Cirloganu \textit{et al.} propose that the electron wavefunctions are spread over the whole structure, while the hole wavefunctions are localised around the core creating the desired spatial separation, and proceed to study QDs with a \ce{PbSe} core and a \ce{CdSe} shell of varying sizes; such dots can be characterised by an aspect ratio $\rho$, defined as the ratio of shell thickness to the overall radius. Thus,  consistently higher MEG rates than reported for \ce{PbSe} QDs were observed for core-shell QDs. 
A critical aspect ratio, which would provide the energy-conservation limit of $\eta = 1$, is predicted to lie in the interval $\rho_c = 0.45-0.7$. Further related analysis on the effect of core-shell aspect ratio, as well as of the degree of alloying between the core and the shell and its spectrographic implications can be found in the review \cite{Klimov2014}.  

In the same year, Eshet \textit{et al.} looked at heterostructured nanorods of \ce{CdSe} and \ce{CdS} \cite{Eshet2016} created by a seeded-growth approach \cite{Carbone2007}.  They analysed both seeded and core-shell structures, obtaining results which comply with the standard behaviour above $\hbar \omega_{th}$:  for the seeded structures, the efficiency $\eta$ increases with the seed size, while for the core/shell structures $\eta$ increases as the core-diameter  decreases.  However, when probing the samples in the non-MEG regime ($ \hbar \omega < \hbar \omega_{th}$), it was found that, unexpectedly, $\eta$ actually decreases with decreasing core size,  which is converse to theoretical prediction as well as mono-composite QD behaviour.  A 2016 paper by Eshet \textit{et al.} furthered the inquiry into how $\hbar \omega_{th}$ could be lowered from previous values of around $3E_g$ by analysing the dynamics of nanorod heterostructures whose composition changes from one to the other end of the rod. Theory suggests that in this class of heterostructures the internal electric field generated from photon pumping would aid MEG. When pumping around the energy-conservation limit of $\hbar \omega = 2E_g$ MEG efficiencies of 10\%-20\% were observed.

\begin{figure}
  \begin{center}
\scalebox{0.5}{\includegraphics{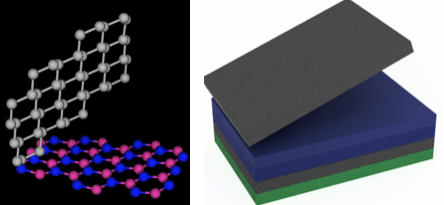}} 
\caption{An example of a graphene Van der Waals structure, using blocks as analogy. Adapted from \cite[A. Geim]{Geim2014}. {\bf Grey} - graphene, {\bf purple} -  \ce{hBN}, {\bf blue} -  \ce{MoS2}, {\bf green} -  \ce{WSe2}.} \label{het2}
  \end{center} 
  \hrulefill
\end{figure}

No semiconductor heterostructures were investigated in 
2D confined systems. However, there has been research into the so-called `Van der Waals' graphene heterostructures \cite{Geim2014} (figure \ref{het2}). These consist of layers of graphene separated by nanosheets of other materials, such as \ce{MoS2} or \ce{WSe2}.  
MEG studies on this kind of `material stacks' proved inconclusive \cite{Wu2016}, and 
further research is needed to shed light as to their potential.  

\section{Summary and Outlook}\label{secsum}

The technology required to create new generations of cheap and efficient solar cells is still a way off its ultimate goal. Despite this, since 2002 there have been significant breakthroughs in the study of nanostructures to overcome the Shockley-Queisser Limit.  Through generating multiple pairs of excitons per incident photon, efficiencies have been measured strongly indicating that this goal should be achievable. As this compact review strived to show, experimental progress has led to a better understanding of the dynamics which govern excitons experiencing 0-, \, 1- and 2-dimensional quantum confinement.  Work by computational and theoretical groups have led to research which allows for better prediction in future candidates for MEG, and better prediction of the effect various factors such as aspect ratio or size might have.  

Quantum Dots in a colloidal environment have now moved on to being studied in device environments. To some extent, monocomposite QDs are being put to one side in favour of studies into heterostructures.   The most state-of-the-art and applicable results have come from the creation of actual devices and measurement high external quantum efficiencies \cite{Zhang2014, Bohm2015, Davis2015}.  Whilst they have yet to show voltages exceeding the Shockley-Queisser Limit, such devices have given external QYs showing MEG does work when turning excitons into carriers. To these aims, the potential of carbon structures is still not well understood in the 1- and 2-dimensional regimes. However, research has conclusively shown that higher-dimensional architectures do reach extremely high efficiencies at threshold frequencies outside the solar spectrum  \cite{Plotzing2014, McClain2010}.  Attempts to reduce the threshold energy by using ribbons, multi-layers of graphene and other systems are underway.

Future investigation into the possibility of creating nanorods and nanosheets devices seem to be in order, as well as into the effect of the nanostructure environment on the MEG efficiency. In particular, a better understanding of the latter seems key to really appreciate the impact these combined advances might have toward potentially revolutionary results. 
It should also be noted that MEG itself is unlikely to be the only pathway towards third generation solar cells. It is one of the many phenomena being explored that could come together in order to create such devices.
It seems fair to predict that, when combined with work into the effect of electric fields on solar voltaic performance, or tandem cells, the application of nanostructures to light harvesting might achieve enhanced efficiencies in both proof of principle experiments and operational devices in the short or medium term.

\section*{Acknowledgements}

The authors acknowledge financial support from EPSRC through grant EP/K026267/1.

{\footnotesize
\bibliographystyle{unsrt}
  \bibliography{LiteratureReview}}

\begin{thebibliography}{10}

\bibitem{Shockley1961}
W.~Shockley and H.~J. Queisser.
\newblock {Detailed balance limit of efficiency of p-n junction solar cells}.
\newblock {\em Journal of Applied Physics}, 32(3):510--519, 1961.

\bibitem{C0JM01178A}
Frederik~C. Krebs, Jan Fyenbo, and Mikkel Jorgensen.
\newblock Product integration of compact roll-to-roll processed polymer solar
  cell modules: methods and manufacture using flexographic printing{,} slot-die
  coating and rotary screen printing.
\newblock {\em J. Mater. Chem.}, 20:8994--9001, 2010.

\bibitem{Scully2010}
M.~O. Scully.
\newblock {Quantum photocell: Using quantum coherence to reduce radiative
  recombination and increase efficiency}.
\newblock {\em Physical Review Letters}, 104(20):1--4, 2010.

\bibitem{Krogstrup2013}
P.~Krogstrup, H.~I. J{\o}rgensen, M.~Heiss, O.~Demichel, J.~V. Holm,
  M.~Aagesen, J.~Nygard, and A.~{Fontcuberta i Morral}.
\newblock {Single-nanowire solar cells beyond the Shockley–Queisser limit}.
\newblock {\em Nature Photonics}, 7(March):1--5, 2013.

\bibitem{Robbins1980}
D.~J. Robbins.
\newblock {Aspects of the Theory of Impact Ionization in Semiconductors (I)}.
\newblock {\em Physica Status Solidi (B)}, 97(1):9--50, 1980.

\bibitem{Badescu2001}
V.~Badescu, P.~T. Landsberg, A.~{De Vos}, and B.~Desoete.
\newblock {Statistical thermodynamic foundation for photovoltaic and
  photothermal conversion. IV. Solar cells with larger-than-unity quantum
  efficiency revisited}.
\newblock {\em Journal of Applied Physics}, 89(4):2482--2490, 2001.

\bibitem{De.Vos1998}
A.~de. Vos and B.~Desoete.
\newblock {On the ideal performance of solar cells with larger-than-unity
  quantum efficiency}.
\newblock {\em Solar Energy Materials and Solar Cells}, 51(3-4):413--424, 1998.

\bibitem{DeVos1993}
A.~{De Vos}, P.~T. Landsberg, P.~Baruch, and J.~E. Parrott.
\newblock {Entropy fluxes, endoreversibility, and solar energy conversion}.
\newblock {\em Journal of Applied Physics}, 74(6):3631--3637, 1993.

\bibitem{Gent1995}
P.~Baruch, A.~{De Vos}, P.~T. Landsberg, and J.E. Parrott.
\newblock {On some thermodynamic aspects of photovoltaic solar energy
  conversion}.
\newblock {\em Solar Energy Materials and Solar Cells}, 36(2):201--222, 1995.

\bibitem{Beard2010}
M.~C. Beard, A.~G. Midgett, M.~C. Hanna, J.~M. Luther, B.~K. Hughes, and A.~J.
  Nozik.
\newblock {Comparing multiple exciton generation in quantum dots to impact
  ionization in bulk semiconductors: Implications for enhancement of solar
  energy conversion}.
\newblock {\em Nano Letters}, 10(8):3019--3027, 2010.

\bibitem{Trupke2002}
T.~Trupke, M.~A. Green, and P.~W{\"{u}}rfel.
\newblock {Improving solar cell efficiencies by down-conversion of high-energy
  photons of high-energy photons}.
\newblock {\em Journal of Applied Physics}, 1668, 2002.

\bibitem{Badescu2007}
V.~Badescu, A.~De~Vos, and A.~M. Badescu.
\newblock {Improved model for solar cells with down-conversion and
  down-shifting of high-energy photons}.
\newblock {\em Journal of Applied Physics}, 2007.

\bibitem{nozik2002quantum}
A.~J. Nozik.
\newblock Quantum dot solar cells.
\newblock {\em Physica E: Low-dimensional Systems and Nanostructures},
  14(1):115--120, 2002.

\bibitem{PhysRevLett.92.186601}
R.~D. Schaller and V.~I. Klimov.
\newblock High efficiency carrier multiplication in pbse nanocrystals:
  Implications for solar energy conversion.
\newblock {\em Phys. Rev. Lett.}, 92:186601, May 2004.

\bibitem{Schaller2006}
R.~D. Schaller, M.~Sykora, J.~M. Pietryga, and V.~I. Klimov.
\newblock {Seven excitons at a cost of one: Redefining the limits for
  conversion efficiency of photons into charge carriers}.
\newblock {\em Nano Letters}, 6(3):424--429, 2006.

\bibitem{Beard2007}
M.~C. Beard, K.~P. Knutsen, P.~Yu, J.~M. Luther, Q.~Song, W.~K. Metzger, R.~J.
  Ellingson, and A.~J. Nozik.
\newblock {Multiple exciton generation in colloidal silicon nanocrystals}.
\newblock {\em Nano Letters}, 7(8):2506--2512, 2007.

\bibitem{Nair2007}
G.~Nair and M.~G. Bawendi.
\newblock {Carrier multiplication yields of CdSe and CdTe nanocrystals by
  transient photoluminescence spectroscopy}.
\newblock {\em Phys. Rev. B}, 76:81304, 2007.

\bibitem{Pijpers2007}
J.~J.~H. Pijpers, E.~Hendry, M.~T.~W. Milder, R.~Fanciulli, J.~Savolainen,
  J.~L. Herek, D.~Vanmaekelbergh, S.~Ruhman, D.~Mocatta, D.~Oron, A.~Aharoni,
  U.~Banin, and M.~Bonn.
\newblock {Carrier multiplication and its reduction by photodoping in colloidal
  InAs quantum dots}.
\newblock {\em Journal of Physical Chemistry C}, 111(11):4146--4152, 2007.

\bibitem{Gabor2009}
N.~M. Gabor, Z.~Zhong, K.~Bosnick, J.~Park, and P.~L. McEuen.
\newblock {Extremely efficient multiple electron-hole pair generation in carbon
  nanotube photodiodes}.
\newblock {\em Science}, 325(5946):1367--1371, 2009.

\bibitem{Gao2015}
J.~Gao, A.~F. Fidler, and V.~I. Klimov.
\newblock {Carrier multiplication detected through transient photocurrent in
  device-grade films of lead selenide quantum dots}.
\newblock {\em Nature Communications}, 6:----, 2015.

\bibitem{Stolle2013}
C.~J. Stolle, T.~B. Harvey, and B.~A. Korgel.
\newblock {Nanocrystal photovoltaics : a review of recent progress}.
\newblock {\em Current Opinion in Chemical Engineering}, pages 1--8, 2013.

\bibitem{Jaeger2013}
H.~M. Jaeger, K.~I.~M. Hyeon-deuk, and O.~V. Prezhdo.
\newblock {Exciton Multiplication from First Principles.pdf}.
\newblock {\em Accounts of Chemical Research}, 46(6), 2013.

\bibitem{Ellingson2005}
R.~Ellingson, M.~Beard, J.~Johnson, P.~Yu, O.~Micic, A.~Nozik, A.~Shabaev, and
  A.~Efros.
\newblock {Highly Efficient Multiple Exciton Generation in Colloidal PbSe and
  PbS Quantum Dots}.
\newblock {\em Nano Letters}, 13(3):1092--1099, 2005.

\bibitem{Beard2011}
M.~C. Beard.
\newblock {Multiple exciton generation in semiconductor quantum dots}.
\newblock {\em Journal of Physical Chemistry Letters}, 2(11):1282--1288, 2011.

\bibitem{doi:10.1021/nl100177c}
J.~A. McGuire, M.~Sykora, J.~Joo, J.~M. Pietryga, and V.~I. Klimov.
\newblock Apparent versus true carrier multiplication yields in semiconductor
  nanocrystals.
\newblock {\em Nano Letters}, 10(6):2049--2057, 2010.
\newblock PMID: 20459066.

\bibitem{Midgett2010}
A.~G. Midgett, H.~W. Hillhouse, B.~K. Hughes, A.~J. Nozik, and M.~C. Beard.
\newblock {Flowing versus Static Conditions for Measuring Multiple Exciton
  Generation in PbSe Quantum Dots}.
\newblock {\em The Journal of Physical Chemistry C}, 114(41):17486--17500,
  2010.

\bibitem{B913277P}
Y.~Kanai, Z.~Wu, and J.~C. Grossman.
\newblock Charge separation in nanoscale photovoltaic materials: recent
  insights from first-principles electronic structure theory.
\newblock {\em J. Mater. Chem.}, 20:1053--1061, 2010.

\bibitem{Kumar2016}
M.~Kumar, S.~Vezzoli, Z.~Wang, V.~Chaudhary, R.~Ramanujan, G.~Gurzadyan,
  A.~Bruno, and C.~Soci.
\newblock {Hot exciton cooling and multiple exciton generation in PbSe quantum
  dots}.
\newblock {\em Phys. Chem. Chem. Phys.}, 18:31107--31114, 2016.

\bibitem{Shabaev2013}
A.~Shabaev, C.~S. Hellberg, and A.~L. Efros.
\newblock {Efficiency of multiexciton generation in colloidal nanostructures}.
\newblock {\em Accounts of Chemical Research}, 46(6):1242--1251, 2013.

\bibitem{Piryatinski2010}
A.~Piryatinski and K.~A. Velizhanin.
\newblock {An exciton scattering model for carrier multiplication in
  semiconductor nanocrystals: Theory}.
\newblock {\em Journal of Chemical Physics}, 133(8), 2010.

\bibitem{LeBris2010}
A.~{Le Bris} and J.~F. Guillemoles.
\newblock {Hot carrier solar cells: Achievable efficiency accounting for heat
  losses in the absorber and through contacts}.
\newblock {\em Applied Physics Letters}, 97(11):2010--2013, 2010.

\bibitem{doi:10.1021/nl062059v}
A.~Shabaev, Al.~L. Efros, and A.~J. Nozik.
\newblock Multiexciton generation by a single photon in nanocrystals.
\newblock {\em Nano Letters}, 6(12):2856--2863, 2006.
\newblock PMID: 17163719.

\bibitem{Piryatinski2007}
A.~Piryatinski, S.~A. Ivanov, S.~Tretiak, and V.~I. Klimov.
\newblock {Effect of Quantum and Dielectric Confinement on the Exciton −
  Exciton Interaction Energy in Type II Core / Shell Semiconductor
  Nanocrystals}.
\newblock {\em Nano Letters}, 7(1):108--115, 2007.

\bibitem{Stewart2013}
J.~T. Stewart, L.~A. Padilha, W.~K. Bae, W.~K. Koh, J.~M. Pietryga, and V.~I.
  Klimov.
\newblock {Carrier multiplication in quantum dots within the framework of two
  competing energy relaxation mechanisms}.
\newblock {\em Journal of Physical Chemistry Letters}, 4(12):2061--2068, 2013.

\bibitem{Lan2014}
X.~Lan, S.~Masala, and E.~H. Sargent.
\newblock {Charge-extraction strategies for colloidal quantum dot
  photovoltaics}.
\newblock {\em Nat Mater}, 13(3):233--240, 2014.

\bibitem{RevModPhys.86.153}
I.~M. Georgescu, S.~Ashhab, and F.~Nori.
\newblock Quantum simulation.
\newblock {\em Rev. Mod. Phys.}, 86:153--185, Mar 2014.

\bibitem{2016APS..MARH24002M}
D.~{Mihaylov}, A.~{Kryjevski}, D.~{Kilin}, S.~{Kilina}, and D.~{Vogel}.
\newblock {Multiple Exciton Generation in Semiconductor Nxanostructures:
  DFT-based Computation}.
\newblock In {\em APS Meeting Abstracts}, 2016.

\bibitem{BEARD2014}
M.~C. Beard, H.~I.~P. Alexander, J.~M. Luther, E.~H. Sargent, and A.~J. Nozik.
\newblock {\em {Quantum Confined Semiconductors for Enhancing Solar
  Photoconversion through Multiple Exciton Generation}}.
\newblock Number~11 in 1. Research Gate, 2014.

\bibitem{1Padilha2013}
Lazaro~A. Padilha, John~T. Stewart, Richard~L. Sandberg, Wan~Ki Bae, Weon~Kyu
  Koh, Jeffrey~M. Pietryga, and Victor~I. Klimov.
\newblock {Carrier multiplication in semiconductor nanocrystals: Influence of
  size, shape, and composition}.
\newblock {\em Accounts of Chemical Research}, 46(6):1261--1269, 2013.

\bibitem{Pijpers2009}
J.~J.~H. Pijpers, R.~Ulbricht, K.~J. Tielrooij, A.~Osherov, Y.~Golan,
  C.~Delerue, G.~Allan, and M.~Bonn.
\newblock {Assessment of carrier-multiplication efficiency in bulk PbSe and
  PbS}.
\newblock {\em Nat Phys}, 5(11):811--814, nov 2009.

\bibitem{Zhang2014}
J.~Zhang, J.~Gao, C.~P. Church, E.~M. Miller, J.~M. Luther, V.~I. Klimov, and
  M.~C. Beard.
\newblock {PbSe Quantum Dot Solar Cells with More than 6{\%} E ffi ciency
  Fabricated in Ambient Atmosphere}.
\newblock {\em Nano Letters}, 2014.

\bibitem{Koole2008}
R.~Koole, G.~Allan, C.~Delerue, A.~Meijerink, D.~Vanmaekelbergh, and A.~J.
  Houtepen.
\newblock {Optical investigation of quantum confinement in PbSe nanocrystals at
  different points in the brillouin zone}.
\newblock {\em Small}, 4(1):127--133, 2008.

\bibitem{Midgett2013}
A.~G. Midgett, J.~M. Luther, J.~T. Stewart, D.~K. Smith, L.~Padilha, V.~I.
  Klimov, A.~J. Nozik, and M.~C. Beard.
\newblock {Size and Composition Dependent Multiple Exciton Generation
  Efficiency in PbS, PbSe, and PbS}.
\newblock {\em Nano letters}, 13(2):3079, 2013.

\bibitem{Stewart2012}
J.~T. Stewart, L.~A. Padilha, M.~M. Qazilbash, J.~M. Pietryga, A.~G. Midgett,
  J.~M. Luther, M.~C. Beard, A.~J. Nozik, and V.~I. Klimov.
\newblock {Comparison of carrier multiplication yields in PbS and PbSe
  nanocrystals: The role of competing energy-loss processes}.
\newblock {\em Nano Letters}, 12(2):622--628, 2012.

\bibitem{Hao2009}
X.~J. Hao, E.~C. Cho, G.~Scardera, Y.~S. Shen, E.~Bellet-Amalric, D.~Bellet,
  G.~Conibeer, and M.~A. Green.
\newblock {Phosphorus-doped silicon quantum dots for all-silicon quantum dot
  tandem solar cells}.
\newblock {\em Solar Energy Materials and Solar Cells}, 93(9):1524--1530, 2009.

\bibitem{Bergren2016}
M.~R. Bergren, P.~K.~B. Palomaki, N.~R. Neale, T.~E. Furtak, and M.~C. Beard.
\newblock {Size-Dependent Exciton Formation Dynamics in Colloidal Silicon
  Quantum Dots}.
\newblock {\em ACS Nano}, 10(2):2316--2323, 2016.

\bibitem{Trinh2012}
M.~T. Trinh, R.~Limpens, W.~D. A.~M. de~Boer, J.~M. Schins, L.~D.~A. Siebbeles,
  and T.~Gregorkiewicz.
\newblock {Direct generation of multiple excitons in adjacent silicon
  nanocrystals revealed by induced absorption}.
\newblock {\em Nature Photonics}, 6(5):316--321, 2012.

\bibitem{Kryjevski2015a}
A.~Kryjevski and D.~Kilin.
\newblock {Enhanced multiple exciton generation in amorphous silicon nanowires
  and films}.
\newblock {\em Molecular Physics}, 8976(September):1--15, 2015.

\bibitem{Saeed2015}
S.~Saeed, C.~{de Weerd}, P.~Stallinga, F.~C.~M. Spoor, A.~J. Houtepen, L.~D.~A.
  Siebbeles, and T.~Gregorkiewicz.
\newblock {Carrier multiplication in germanium nanocrystals}.
\newblock {\em Light: Science {\&} Applications}, 4(2):e251, 2015.

\bibitem{Bohm2015}
M.~L. Bohm, T.~C. Jellicoe, M.~Tabachnyk, F.~Davis, N. J. L. K.and
  Wisnivesky-Rocca-Rivarola, C.~Ducati, B.~Ehrler, A.~A. Bakulin, and N.~C.
  Greenham.
\newblock {Lead telluride quantum dot solar cells displaying external quantum
  efficiencies exceeding 120{\%}}.
\newblock {\em Nano Letters}, 15(12):7987--7993, 2015.

\bibitem{Bartnik2010}
A.~C. Bartnik, A.~L. Efros, W.~K. Koh, C.~B. Murray, and F.~W. Wise.
\newblock {Electronic states and optical properties of PbSe nanorods and
  nanowires}.
\newblock {\em Physical Review B - Condensed Matter and Materials Physics},
  82(19):1--16, 2010.

\bibitem{Li2001}
L.~S. Li, J.~Hu, W.~Yang, and A.~P. Alivisatos.
\newblock {Band Gap Variation of Size- and Shape-Controlled Colloidal CdSe
  Quantum Rods}.
\newblock {\em Nano Letters}, 1(7):349--351, 2001.

\bibitem{Kryjevski2015}
A.~Kryjevski and D.~Kilin.
\newblock {Enhanced multiple exciton generation in amorphous silicon nanowires
  and films}.
\newblock {\em Molecular Physics}, 8976(September):1--15, 2015.

\bibitem{Garnett2011}
Erik~C Garnett, Mark~L Brongersma, Yi~Cui, and Michael~D Mcgehee.
\newblock {Nanowire Solar Cells}.
\newblock {\em Annual Reviews}, 41:269--295, 2011.

\bibitem{Padilha2013}
L.~A. Padilha, J.~T. Stewart, R.~L. Sandberg, W.~K. Bae, W.~K. Koh, J.~M.
  Pietryga, and V.~I. Klimov.
\newblock {Aspect ratio dependence of auger recombination and carrier
  multiplication in PbSe nanorods}.
\newblock {\em Nano Letters}, 13(3):1092--1099, 2013.

\bibitem{McGuire2008}
J.~A. McGuire.
\newblock {New aspects of carrier multiplication in semiconductor
  nanocrystals}.
\newblock {\em Acc. Chem. Res.}, 41:1810--1819, 2008.

\bibitem{McGuire2010}
J.~A. McGuire, M.~Sykora, J.~Joo, J.~M. Pietryga, and V.~I. Klimov.
\newblock {Apparent versus true carrier multiplication yields in semiconductor
  nanocrystals}.
\newblock {\em Nano Letters}, 10(6):2049--2057, 2010.

\bibitem{koh2010}
W.~K. Koh, A.~C. Bartnik, F.~W. Wise, and C.r~B. Murray.
\newblock {Synthesis of monodisperse PbSe nanorods: A case for oriented
  attachment}.
\newblock {\em Journal of the American Chemical Society}, 132(11):3909--3913,
  2010.

\bibitem{Wang2010}
H.~Wang, Y.~Liu, M.~Li, H.~Huang, H.~M. Xu, R.~J. Hong, and H.~Shen.
\newblock {Facile synthesis of ultra-small PbSe nanorods for photovoltaic
  application}.
\newblock {\em Optoelectronics and Advanced Materials, Rapid Communications},
  4(8):1166--1169, 2010.

\bibitem{Sandberg2012}
R.~L. Sandberg, L.~A. Padilha, M.~M. Qazilbash, W.~K. Bae, R.~D. Schaller,
  J.~M. Pietryga, M.~J. Stevens, B.~Baek, S.~W. Nam, and V.~I. Klimov.
\newblock {Multiexciton dynamics in infrared-emitting colloidal nanostructures
  probed by a superconducting nanowire single-photon detector}.
\newblock {\em ACS Nano}, 6(11):9532--9540, 2012.

\bibitem{Cunningham2011}
Paul~D. Cunningham, Janice~E. Boercker, Edward~E. Foos, Matthew~P. Lumb, A.~R.
  Smith, J.~G. Tischler, and J.~S. Melinger.
\newblock {Enhanced multiple exciton generation in quasi-one-dimensional
  semiconductors}.
\newblock {\em Nano Letters}, 11(8):3476--3481, 2011.

\bibitem{Aerts2014}
M.~Aerts, T.~Bielewicz, C.~Klinke, F.~C. Grozema, A.~J. Houtepen, J.~M. Schins,
  and Laurens.~D. Siebbeles.
\newblock {Highly efficient carrier multiplication in PbS nanosheets.}
\newblock {\em Nature communications}, 5:3789, 2014.

\bibitem{Davis2015}
N.~J.L.K. Davis, M.~L. Bo, M.~Tabachnyk, F.~Wisnivesky-Rocca-Rivarola, C.~J.
  Tom, D.~Caterina, B.~Ehrler, and N.~Greenham.
\newblock {Multiple-exciton generation in lead selenide nanorod solar cells
  with external quantum efficiencies exceeding 120{\%}}.
\newblock {\em Nature Communications}, 6(2):137--141, 2015.

\bibitem{1Cunningham2013}
P.~Cunningham, J.~Boercker, E.~Foos, M.~Lumb, A.~Smith, J.~Tischler, and
  J.~Melinger.
\newblock {Correction to Enhanced Multiple Exciton Generation in Quasi-One-
  Dimensional Semiconductors}.
\newblock {\em Nano Letters}, 6(11):401525, 2013.

\bibitem{Spataru2004}
C.~Spataru, S.~Ismail-Beigi, L.~Benedict, and S.~Louie.
\newblock {Excitonic effects and optical spectra of single-walled carbon
  nanotubes}.
\newblock {\em Physical Review Letters}, 92(7):077402, 2004.

\bibitem{Perebeinos2004}
V.~Perebeinos, J.~T., and P.~Avouris.
\newblock {Scaling of excitons in carbon nanotubes}.
\newblock {\em Physical Review Letters}, 92(25 I):257402--1, 2004.

\bibitem{Ueda2008}
K.~Ueda, A.and~Matsuda, T.~Tayagaki, and Y.~Kanemitsu.
\newblock {Carrier multiplication in carbon nanotubes studied by femtosecond
  pump-probe spectroscopy}.
\newblock {\em Applied Physics Letters}, 92(23), 2008.

\bibitem{Wang2010a}
S.~Wang, M.~Khafizov, X.~Tu, M.~Zheng, and T.~D. Krauss.
\newblock {Multiple exciton generation in single-walled carbon nanotubes}.
\newblock {\em Nano Letters}, 10(7):2381--2386, 2010.

\bibitem{Yuma2013}
B.~Yuma, S.~Berciaud, J.~Besbas, J.~Shaver, S.~Santos, S.~Ghosh, R.~B. Weisman,
  L.~Cognet, M.~Gallart, M.~Ziegler, B.~H{\"{o}}nerlage, B.~Lounis, and
  P.~Gilliot.
\newblock {Biexciton, single carrier, and trion generation dynamics in
  single-walled carbon nanotubes}.
\newblock {\em Physical Review B - Condensed Matter and Materials Physics},
  87(20):1--7, 2013.

\bibitem{Tune2013}
D.~D. Tune and J.~G. Shapter.
\newblock {The potential sunlight harvesting efficiency of carbon nanotube
  solar cells}.
\newblock {\em Energy {\&} Environmental Science}, 6(9):2572--2577, 2013.

\bibitem{Jariwala2013}
D.~Jariwala, V.~K. Sangwan, L.~J. Lauhon, T.~J. Marks, and M.~C. Hersam.
\newblock {Carbon nanomaterials for electronics, optoelectronics,
  photovoltaics, and sensing.}
\newblock {\em Chemical Society reviews}, 42:2824--60, 2013.

\bibitem{Soavi2016}
G.~Soavi, S.~{Dal Conte}, C.~Manzoni, D.~Viola, A.~Narita, Y.~Hu, X.~Feng,
  U.~Hohenester, E.~Molinari, D.~Prezzi, K.~M{\"{u}}llen, and G.~Cerullo.
\newblock {Exciton-exciton annihilation and biexciton stimulated emission in
  graphene nanoribbons.}
\newblock {\em Nature communications}, 7:11010, 2016.

\bibitem{Cai2010}
J.~Cai, P.~Ruffieux, R.~Jaafar, M.~Bieri, T.~Braun, S.~Blankenburg, M.~Muoth,
  A.~P. Seitsonen, M.~Saleh, X.~Feng, K.~Mullen, and R.~Fasel.
\newblock {Atomically precise bottom-up fabrication of graphene nanoribbons}.
\newblock {\em Nature}, 466(7305):470--473, 2010.

\bibitem{Prezzi2011}
D.~Prezzi, A.~Varsano, D.and~Ruini, and E.~Molinari.
\newblock {Quantum dot states and optical excitations of edge-modulated
  graphene nanoribbons}.
\newblock {\em Physical Review B - Condensed Matter and Materials Physics},
  84(4):1--4, 2011.

\bibitem{Yang2007}
L.~Yang, M.~L. Cohen, and S.~G. Louie.
\newblock {Excitonic Effects in the Optical Spectra of Graphene Nanoribbons}.
\newblock {\em Nano Lett.}, 7(10):3112--3115, 2007.

\bibitem{Prezzi2007}
D.~Prezzi, D.~Varsano, A.~Ruini, A.~Marini, and E.~Molinari.
\newblock {Optical properties of graphene nanoribbons: The role of many-body
  effects}.
\newblock {\em Physical Review B}, 77(4):41404, 2007.

\bibitem{Nasilowski2016}
M.~Nasilowski, B.~Mahler, E.~Lhuillier, S.~Ithurria, and B.~Dubertret.
\newblock {Two-Dimensional Colloidal Nanocrystals}.
\newblock {\em Chemical Reviews}, 116(18):10934--10982, 2016.

\bibitem{Schliehe2010}
C.~Schliehe, B.~H. Juarez, M.~Pelletier, S.~Jander, D.~Greshnykh, M.~Nagel,
  A.~Meyer, S.~Foerster, A.~Kornowski, C.~Klinke, and H.~Weller.
\newblock {Ultrathin PbS sheets by two-dimensional oriented attachment.}
\newblock {\em Science (New York, N.Y.)}, 329(2010):550--553, 2010.

\bibitem{Bonaccorso2010}
F.~Bonaccorso, Z.~Sun, T.~Hasan, and A.~C. Ferrari.
\newblock {Graphene Photonics and Optoelectronics}.
\newblock {\em Nature Photonics}, 4(9):611--622, 2010.

\bibitem{Winzer2010}
T.~Winzer, A.~Knorr, and E.~Malic.
\newblock {Carrier multiplication in graphene}.
\newblock {\em Nano Letters}, 10(12):4839--4843, 2010.

\bibitem{Winzer2012}
T.~Winzer and E.~Mali{\'{c}}.
\newblock {Impact of Auger processes on carrier dynamics in graphene}.
\newblock {\em Physical Review B - Condensed Matter and Materials Physics},
  85(24):1--5, 2012.

\bibitem{Tielrooij2013}
K.~J. Tielrooij, J.~C.~W. Song, S.~A. Jensen, A.~Centeno, A.~Pesquera, A.~Z.
  Elorza, M.~Bonn, L.~S. Levitov, and F.~H~L Koppens.
\newblock {Photoexcitation cascade and multiple hot carrier generation in
  graphene}.
\newblock {\em 2013 Conference on Lasers and Electro-Optics Europe and
  International Quantum Electronics Conference, CLEO/Europe-IQEC 2013},
  9(4):1--5, 2013.

\bibitem{Plotzing2014}
T.~Pl{\"{o}}tzing, T.~Winzer, E.~Malic, D.~Neumaier, A.~Knorr, and H.~Kurz.
\newblock {Experimental verification of carrier multiplication in graphene}.
\newblock {\em Nano Letters}, 14(9):5371--5375, 2014.

\bibitem{Ubani2016}
C.~A. Ubani, M.~A. Ibrahim, M.~A.~M. Teridi, K.~Sopian, J.~Ali, and K.~T.
  Chaudhary.
\newblock {Application of graphene in dye and quantum dots sensitized solar
  cell}.
\newblock {\em Solar Energy}, 137:531--550, 2016.

\bibitem{Cirloganu2014}
C.~M. Cirloganu, L.~A. Padilha, Q.~Lin, N.~S. Makarov, K.~Velizhanin, H.~Luo,
  I.~Robel, J.~M. Pietryga, and V.~I. Klimov.
\newblock {Enhanced carrier multiplication in engineered quasi-type-II quantum
  dots}.
\newblock {\em Nature Communications}, 5(May):4148, 2014.

\bibitem{Klimov2014}
V.~I. Klimov.
\newblock {Multicarrier Interactions in Semiconductor Nanocrystals in Relation
  to the Phenomena of Auger Recombination and Carrier Multiplication}.
\newblock {\em Annual Review of Condensed Matter Physics}, 5(1):285--316, 2014.

\bibitem{Eshet2016}
H.~Eshet, R.~Baer, D.~Neuhauser, and E.~Rabani.
\newblock {Theory of highly efficient multiexciton generation in type-II
  nanorods}.
\newblock {\em Nature Communications}, 7(May):1--6, 2016.

\bibitem{Carbone2007}
L.~Carbone, C.~Nobile, M.~{De Giorgi}, F.~{Della Sala}, G.~Morello, P.~Pompa,
  M.~Hytch, E.~Snoeck, A.~Fiore, I.~R. Franchini, M.~Nadasan, A.~F. Silvestre,
  L.~Chiodo, S.~Kudera, and L.~Manna.
\newblock {Synthesis and micrometer-scale assembly of colloidal CdSe / CdS
  nanorods prepared by a seeded growth approach}.
\newblock {\em Nano Letters}, 7(10):2942--2950, 2007.

\bibitem{Geim2014}
A.~K. Geim and I.~V. Grigorieva.
\newblock {Van der Waals heterostructures}.
\newblock {\em Nature}, 499(7459):419--425, 2014.

\bibitem{Wu2016}
S.~Wu, L.~Wang, Y.~Lai, W.~Shan, G.~Aivazian, X.~Zhang, T.~Taniguchi,
  K.~Watanabe, D.~Xiao, C.~Dean, J.~Hone, Z.~Li, and X.~Xu.
\newblock {Multiple hot-carrier collection in photo-excited graphene
  Moir{\'{e}} superlattices}.
\newblock {\em Phys. Chem. Letters}, 1(May):1--7, 2016.

\bibitem{McClain2010}
J.~McClain and J.~Schrier.
\newblock {Multiple exciton generation in graphene nanostructures}.
\newblock {\em Journal of Physical Chemistry C}, 114(34):14332--14338, 2010.

\end{thebibliography}

\end{document}